# Artificial Intelligence in Management Studies (2021–2025): A Bibliometric Mapping of Themes, Trends, and Global Contributions

**Yassine SEKAKI,** *(Associate professor)*

*Center for Doctoral Studies: Management, Finance, Digitalization and Applies Statistics*
*Faculty of Legal, Economic and Social Sciences of Tétouan*
*Abdelmalek Essaadi University, Morocco*

**Hamza ZIANE**, *(PhD candidate)*

*Center for Doctoral Studies: Management, Finance, Digitalization and Applies Statistics*
*Faculty of Legal, Economic and Social Sciences of Tétouan*
*Abdelmalek Essaadi University, Morocco*

**Abdelhafid KHAZZAR**, *(PhD candidate)*

*Faculty of Legal, Economic and Social Sciences of Tétouan*
*Abdelmalek Essaadi University, Morocco*

# Artificial Intelligence in Management Studies (2021–2025): A Bibliometric Mapping of Themes, Trends, and Global Contributions

**Abstract**

AI has risen as one of the most influential topics in scientific research over the past decade, focusing on the application of AI technologies in multiple disciplines. Management, as a science that has been studied, researched, and applied, shows great opportunities in terms of AI applications, notably as a tool for enhancing decision-making, sustainability in managerial practices, and so on.

This bibliometric study aims to investigate the trends of AI research in management, defining the focus areas and research gap publications were gathered from the scopus database, and were processed through the bibliometrix library of the R studio analysis software , data were represented using the VOSviewer software,the research yielded a significant number of papers on different topics, data analysis comprised the country of origin, publication journal, author and keyword co-occurrence.

The results showed an increase in publications stuying the uses of AI in management from 2021 to 2024 and a sharp decline of publication output in 2025. leading countries were shown to be China, india and the United States which dominated in volume of publication, while the United Kingdom dominated in citation impact. The dominating journals were ustainability, Technological Forecasting and Social Change, and IEEE Transactions on Engineering Management respectively. The Keywords co-occurrence and thematic mapping showed an increasing shift from technical applications to societal aspects such as sustainability, digital transformation, and decision-making support.

The research shows a strategic shift in AI research, a changing landscape of pure and applied AI toward a more nuanced type of studied.

**Classification JEL**: O32
**Paper type**: Theoretical Research.

## 1. Introduction

The ability of machines to execute human intelligence tasks like learning and reasoning and problem-solving and language understanding defines Artificial Intelligence (AI) which drives substantial changes in modern science and industry according to Mariani & Borghi (2021). The technology includes multiple capabilities which range from basic data collection and management to advanced predictive analytics and natural language processing and image recognition and autonomous decision-making (Russell & Norvig, 2020).

Scientific research has thoroughly studied AI implementation throughout different fields with special focus on its effects in applied sciences and informatics and engineering and economics and social sciences (Brynjolfsson & McAfee, 2017). The field of management stands out as a crucial area of research because AI functions both as an operational tool and as a strategic transformation enabler (Davenport & Ronanki, 2018).

Management, traditionally defined as the coordinated activities of planning, organizing, leading, and controlling organizational resources to achieve specific goals (Robbins & Coulter, 2018), is inherently data-intensive. As organizations navigate environments characterized by complexity, uncertainty, and fast-paced change, the capacity to collect, process, and interpret data becomes crucial. In this context, data-driven decision-making has gained prominence, emphasizing the systematic use of empirical data to support managerial choices (Provost & Fawcett, 2013).

Building on this foundation, AI-assisted decision-making represents the next frontier, where intelligent algorithms analyze large volumes of structured and unstructured data to generate insights, predict trends, assess risks, and recommend actions (Mariani & Borghi, 2021). This not only improves decision speed and accuracy but also reduces cognitive overload, allowing managers to focus on strategic thinking and human-centered leadership.

AI has profoundly transformed multiple facets of management, delivering enhanced capabilities across strategic, operational, organizational, and governance dimensions. At the strategic level, AI-driven tools such as predictive analytics and scenario modeling enable leaders to anticipate potential future outcomes with greater accuracy and to adjust policies proactively. By leveraging large datasets and sophisticated algorithms, decision-makers can simulate diverse scenarios, evaluate risks, and optimize strategic planning processes, ultimately improving organizational agility and long-term competitiveness (Chatterjee et al., 2021).

Artificial Intelligence drives operational management through automation and optimization of essential business functions including supply chain management and human resources and finance and customer relationship management (Liengpunsakul, 2021). Through automation systems organizations achieve better operational performance because automation reduces manual errors while AI optimization models enhance resource allocation efficiency and streamline processes. The combination of automated operations with AI optimization leads to improved performance while providing organizations real-time responses to market changes and customer requirements (Bughin et al., 2018).

Through AI organizations learn better and innovate more because AI enables knowledge management systems to collect and analyze information across their entire enterprise. The combination of these systems creates a data exploration environment that supports employee experimentation with evidence-based problem-solving for developing new innovative approaches. The dynamic operational environment enables the creation of innovative products and business models which leads to long-term competitive advantage (Wamba-Taguimdje et al., 2020).

Risk management and governance frameworks experience significant improvement because AI provides advanced anomaly detection as well as regulatory compliance monitoring along with decision support in complex high-stakes environments (Wamba-Taguimdje et al., 2020).

Through pattern recognition and real-time analysis AI systems detect irregularities and potential threats early so organizations can better protect themselves from risks. AI-based governance tools enhance transparency while providing accountability functions and informed oversight capabilities to maintain legal and ethical compliance in organizational practices (Dwivedi et al., 2021).

The adoption of AI in scientific research has expanded dramatically throughout the last twenty years because experts have increasingly recognized AI's transformative abilities across different fields. Research on AI during the late 20th century concentrated on developing fundamental algorithms and theoretical frameworks (Russell & Norvig, 2016). The development of applied AI research accelerated exponentially since the mid-2000s because of enhanced computational capabilities and better machine learning methods and increased availability of data (LeCun et al., 2015).

The adoption of AI research in management studies began later but gained momentum during the early part of the 2010s. The first investigations focused on how AI could automate routine managerial responsibilities as well as help with decision-making functions (Brynjolfsson & McAfee, 2014). The following decade showed expanded usage of AI applications which penetrated strategic planning and human resource management and supply chain optimization and customer relationship management among other domains (Wamba-Taguimdje et al., 2020).

The bibliometric studies of AI in management literature act as essential tools for establishing a systematic overview of the fast-evolving research domain (Donthu et al., 2021). Bibliometric analysis of scientific literature through quantitative examination of publications and citations along with keywords and co-authorship networks reveals major themes and important researchers and institutions while uncovering new trends and research deficiencies (Aria & Cuccurullo, 2017). The analysis enables a complete comprehension of how management researchers interpret and operationalize AI concepts along with their relationships.

The objectives of bibliometric research on AI management literature consist of:

- A chronological analysis of AI research in management reveals periods of growth or stagnation or thematic evolution.
- The research establishes thematic clusters which show major investigation areas including AI-supported decision processes and operational automation and ethical considerations.
- The evaluation reveals the main academic publications along with prominent writers who contribute to knowledge transfer and academic direction.
- The analysis reveals upcoming research directions through the identification of new emerging trends and niche topics.
- The assessment investigates how AI affects management across different disciplines and sectors.

The method demonstrates its usefulness according to García-Murillo and Annabi (2002) in their work on technology management and subsequent studies about AI in business (Huang et al., 2020). The speedy growth of AI applications requires ongoing bibliometric evaluation to monitor scientific developments which support future research direction (Jia et al., 2021).

This research conducts a detailed bibliometric evaluation of management AI literature spanning from 2010 through 2025 to reveal developmental patterns and thematic domains and new directions that will influence AI-based management strategies.

## 2. Theoretical background

### 2.1. Traditional Management

Traditional management has long been the backbone of organizational structure, performance, and effectiveness. This model is primarily built upon the principles of authority, hierarchy, and task guidance (Nuhu et al., 2023).

At the core of traditional management lie the four fundamental functions—planning, organizing, leading, and controlling—which serve to achieve organizational objectives. These functions guide managers in their daily responsibilities and offer a systematic approach to managing both people and resources. Planning involves setting specific, measurable, and time-bound goals that are rooted in organizational needs (Boukssessa, 2025). As such, managers are expected to forecast future trends and identify potential threats to develop appropriate strategies.

Organizing entails the allocation of resources, the definition of roles, the establishment of departments and structures, and the clear assignment of responsibilities across different organizational actors (Robbins & Coulter, 2018).

Leadership, in the context of traditional management, is a directive function. Managers, who occupy leadership roles, are perceived as authoritative figures responsible for guiding and motivating employees while ensuring compliance with rules and protocols. Through this function, managers work to ensure that organizational goals are pursued with discipline and consistency. However, this view of leadership diverges significantly from modern approaches that emphasize employee autonomy, creativity, and participatory decision-making (Akın, 2022).

The final function, controlling, focuses on monitoring performance, comparing outcomes against pre-established benchmarks, and making necessary adjustments. Managers use tools such as performance appraisals, budgets, and audits to maintain order and drive efficiency (Maghribi, 2025).

Traditional management relies heavily on stability, predictability, and productivity, aiming to minimize risks, standardize operations, and align outcomes with predefined expectations (Akın, 2022). This approach is largely influenced by Taylor's scientific management theory, which emphasizes the optimization of work methods, and Fayol's administrative theory, which focuses on administrative processes and principles such as unity of command and division of labor (Robbins & Coulter, 2018).

Although this rigid model has proven effective in many contexts—particularly in industries where consistency, risk mitigation, and compliance are essential—it has been found to be less effective in creative or dynamic environments. Critics argue that traditional management often ignores the emotional and psychological dimensions of human work, treating human resources like machines with defined variables rather than as complex individuals influenced by social and emotional factors.

Nevertheless, it is important to recognize that traditional management remains functional in many settings. Often, it is combined with contemporary models to create a more adaptive and performance-oriented management approach (Nuhu et al., 2023).

### 2.2. AI in Management

Artificial Intelligence has become an increasingly important part of managerial sciences in recent years. Its integration into organizational theory aims to optimize decision-making processes, resource allocation, and operational efficiency (Boukssessa, 2025).

As a scientific discipline, AI is rooted in subfields such as machine learning, deep learning, computer science, and natural language processing. Technically, it leverages algorithmic modeling and statistical inference to detect latent and otherwise invisible patterns in large-scale, multidimensional datasets. This enables predictive and prescriptive analytics and supports data-driven decision-making (Mariani & Borghi, 2021).

In the managerial context, AI facilitates data-driven, highly predictive, and prescriptive decision-making, along with real-time optimization. Its applications range from human resource allocation to market segmentation and consumer profiling (Bughin et al., 2018).

Practically, AI is increasingly embedded in various functional domains of management. In human resource management, it is used in natural language processing, applicant tracking systems, automated résumé parsing, and candidate scoring based on semantic fit. It also supports the evaluation of employee engagement, organizational climate, and other performance-related metrics (Chatterjee et al., 2021).

In financial management, AI enables the automation of transactional tasks such as invoice processing and fraud detection. In marketing, it enhances collaborative filtering, optimizes customer lifetime value, and improves overall strategic decision-making (Bughin et al., 2018). Despite its advantages, the integration of AI into management raises several epistemological and ethical concerns, particularly regarding bias in training datasets, the need for workforce upskilling, workflow redesign, and the development of AI-literate leadership. AI thus represents a significant shift in management, combining modern computational capabilities with traditional managerial processes (Brynjolfsson & McAfee, 2017).

The scientific output related to AI has surged over the past decade, particularly in its applications to managerial fields. AI must therefore be used judiciously, supported by robust and responsible research.

## 3. Material and Methods

The research applies bibliometric analysis to track scientific development between Artificial Intelligence (AI) and management from 2021 through 2025. The research team obtained its data from Scopus which stands as a prominent multidisciplinary database that provides comprehensive coverage through its strict indexing standards.

- *Search Strategy*

The search query used Boolean operators to retrieve documents that included "artificial" and "intelligence" and "management" in their titles or abstracts or keywords.

The search query used *"artificial" AND "intelligence" AND "management"* in the TITLE-ABS-KEY field with *PUBYEAR > 2020 AND PUBYEAR < 2026* parameters.

This search yielded 40,835 documents.

- *Stepwise Refinement Process*

The research used multiple stages of filtering to achieve dataset refinement which met the study requirements.

**1- Subject Area Filtering**
The study retained documents from Social Sciences, Business, Management and Accounting, and Economics, Econometrics and Finance categories (n = 10,093).
The research focuses on AI applications for management and economic decision-making and organizational contexts so this selection matches the study's objectives while excluding technical and medical domains.

**2- Document Type Filtering**
The research included articles and conference papers (n = 5,897).
The study retained articles and conference papers because these formats undergo peer review and contain complete research findings whereas book chapters and editorials and technical notes do not.

**3- Language Filtering**
The dataset contained only English publications which totaled 5,624 documents.
English stands as the primary academic publishing language worldwide because it provides maximum accessibility and enables better comparison between research findings.

- *Final Dataset and Analysis*

The final dataset consisted of 5,624 documents. The Bibliometrix package (Aria & Cuccurullo, 2017) in RStudio allowed the analysis of these documents through both quantitative and qualitative methods. The analysis generated bibliometric indicators that included:
- Annual scientific production
- Most influential authors, sources, and institutions
- Co-authorship and co-citation networks
- The analysis used keyword frequency and thematic mapping to detect new research directions.

*Figure 1: Diagram of the Selection Process*

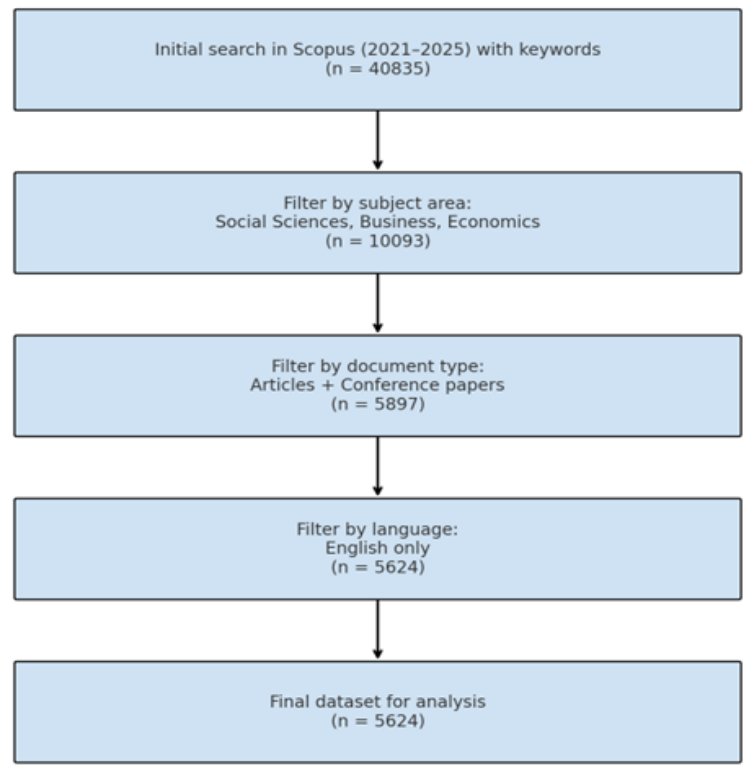

*Source: Created by authors*

## 4. Results

This section presents the main findings of our bibliometric analysis, offering a clear picture of how research on artificial intelligence (AI) in management has evolved between 2021 and 2025. Using quantitative indicators such as yearly publication output, leading journals, top contributing authors and countries, as well as keyword and co-authorship networks, we explore the structure and dynamics of this growing field.

Our results reveal a strong upward trend in scientific output from 2021 to 2024, followed by a noticeable decline in 2025. This slowdown may suggest a shift in research priorities or a consolidation phase after rapid growth. In terms of geographic contribution, China, India, and the United States emerges as the most prolific countries, while the United Kingdom stands out for the impact of its publications, as reflected in citation metrics.

Beyond quantity, the thematic analysis highlights how the field is diversifying. Research is gradually moving from purely technical implementations of AI to broader concerns such as sustainability, digital transformation, and decision-making. These evolving themes point to a more integrated view of AI's role in management, with implications that extend beyond efficiency gains to include strategic and societal considerations.

The following subsections provide a detailed breakdown of these results, starting with an overview of the annual scientific production.

### 4.1. Annual scientific production

Figure 1 displays the trend in scientific production from 2021 to 2025. The period from 2021 to 2022 showed a relative increase in annual production going from 750 paper to slightly less than 1000 paper.

The slope of the trend shifts to a more upward line signifying an increase in overall annual production in the period from 2022 to 2023. The more productive year was 2023-2024 with a very important increase in publication, surpassing the 2000 publication. However, this ascending trajectory reverses in 2024-2025, this period had an exponential decline in annual production, dropping to less than 500 publications.

*Figure 2: Annual scientific production.*

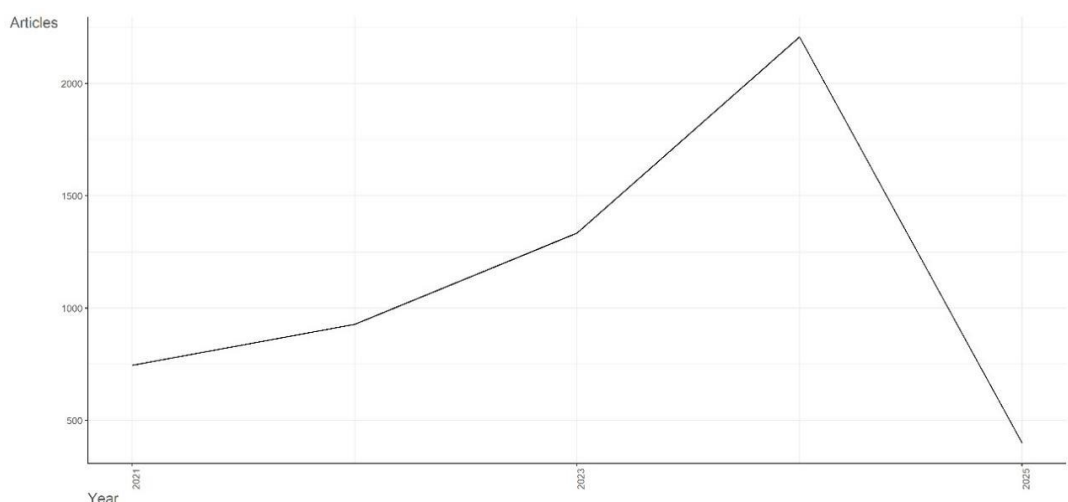

*Source: Created by authors*

### 4.2. Most relevant sources

The temporal variation of sources constitutes an important index in bibliometric analysis, allowing for in depth visualization and analysis of the research trends overtime. Figure 2 shows the temporal variations of the 5 most important sources identified through source identification. Sustainability (Switzerland) shows an exponential increase over the period of 2021-2024 with a drop in the cumulate occurrences trend starting 2024. However, it is important to note that this drop in the trend of cumulate occurrence still shows an increase from 2024 to 2025 but not as high as in the previous period.

The other 5 most relevant sources; water (Switzerland), technological forecasting and social change, studies in system decision and control, journal of cleaner production and IEEE transaction in engineering management; show an overall increase in the cumulate occurrences from 2021-2025, but not as important as in sustainability.

*Figure 3: Sources over time.*

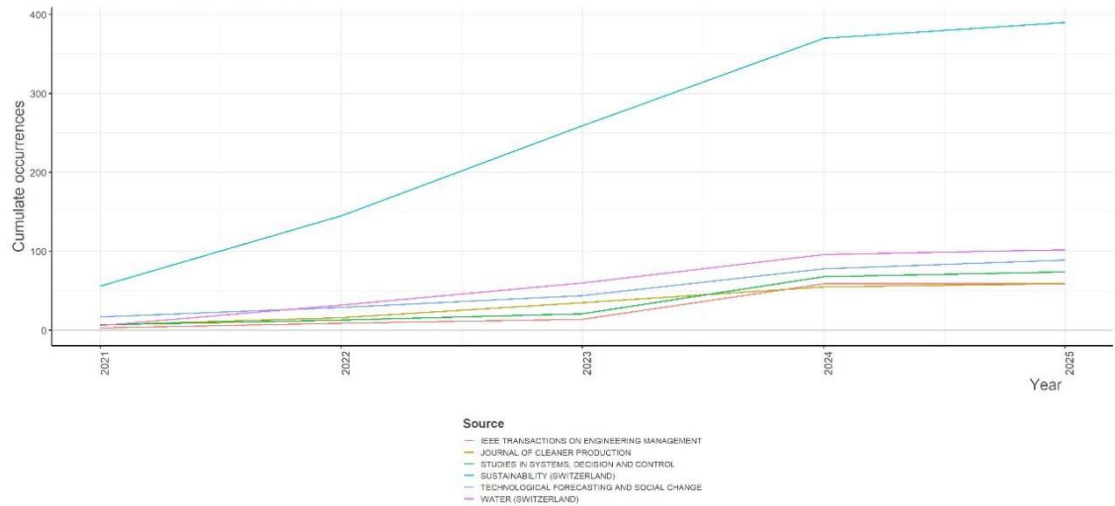

*Source: Created by authors*

### 4.3. Authors' production over time

Defining authors and their contribution to the scientific field is an important bibliometric parameter to be taken into consideration, permitting therefore to acknowledge researchers with a significant impact on the field.

Figure 3 displays authors that have a strong impact on the overall trend of publication in the study period, taking into consideration the number of articles published, and the total citations per year. Wang-y was shown to be more prominent in terms of total articles and citations, followed by li J, li X, chen Y as well as others (Figure 3).

*Figure 4: Author production.*

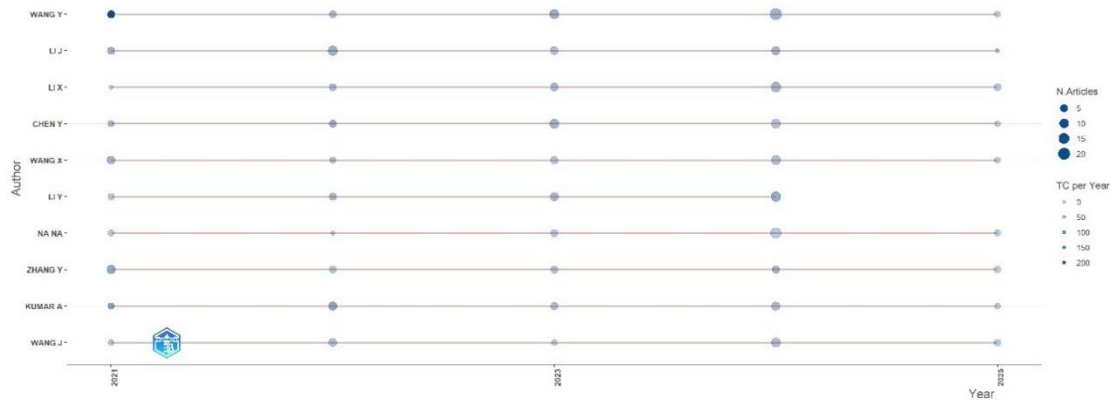

*Source: Created by authors*

### 4.4. Country scientific production

The Figure 4 showed the world map of scientific production by country. It illustrates the geographical disparities in the contribution made to scientific research. The shades of blue go from light to dark, indicating the levels of scientific production from low to high productivity. The countries that showed the highest level of scientific productivity are the United States, China, and India, represented by dark blue color. Japan, Western Europe, Australia, and Canada also showed a high level of scientific activity. Different regions in Latin America, Africa, and Central Asia are lightly colored or grayed out, reflecting a lesser or non-existent contribution in this area.

This global inequality in scientific production is often correlated with the economic development levels, or the investment in the education and innovation, and the support of research of the national policies.

*Figure 5: Country scientific production.*

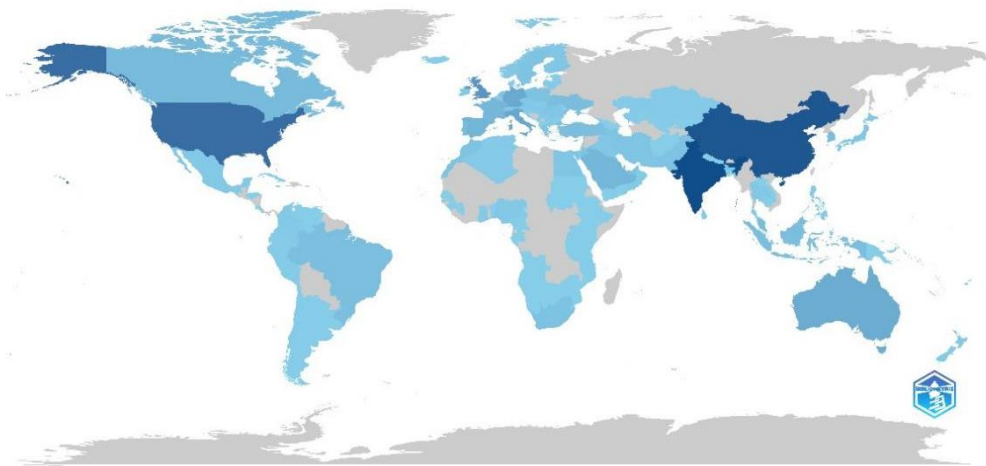

*Source: Created by authors*

### 4.5. Corresponding author's countries

Figure 5 highlights the trends in corresponding author's countries, showing both multi-countries studies (MCP) and single country studies (SCP) affiliations in conjunction with the country and type of study.

*Figure 6: Corresponding authors' countries.*

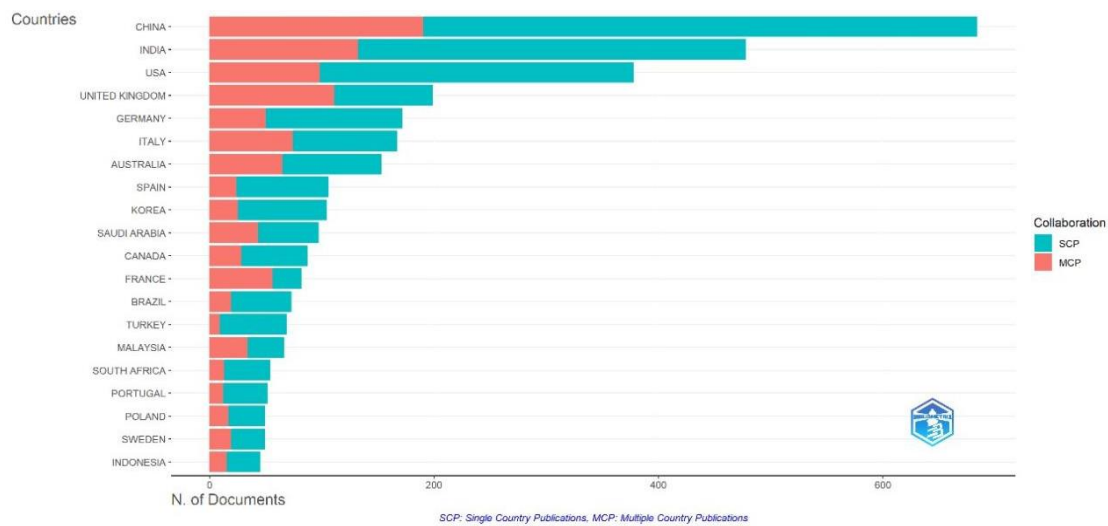

*Source: Created by authors*

These parameters are valuable in bibliometric analysis, as they allowed the establishment of the leaders in the subject at hand both in terms of country and university. Moreover, this type of analysis plays a role in determining the room for collaboration in a specific area.

It is apparent that China leads the way in terms of subject-oriented research, both in SCP's (N>500) and MCP's (N>150) advancing significantly. In second place, India is shown to be quite active in terms of subject driven research showing close to 500 publications and an important number of MCPs close to 150.

In the third place, the USA is also shown to have a significant number of publications on the subject of closing on a total of close to 400 publications also showing good levels of

71

international collaboration surpassing the 150-publication landmark. Other countries such as the UK, Germany, Italy, Australia and others are also listed in Figure 3.

### 4.6. Most Cited Countries

While figure 4 showed that China and India were respectively numbers one and two in terms of the number of papers related to the subject, the number of citations per year shows that the UK, despite being the fourth in the corresponding author's country list, is more influential. It holds the top place in the most cited countries, followed in second place in China and the USA, which were originally ranked first and third in terms of corresponding author countries.

India detains the fourth place in terms of citations indicating that while a lot of papers were published in these countries, the UK, China and the USA's influence on subsequent research is more pronounced. The rest of the most cited countries is constituted by France, Italy, Australia, Finland, and Saudi Arabia.

*Figure 7: Countries ranking by citations.*

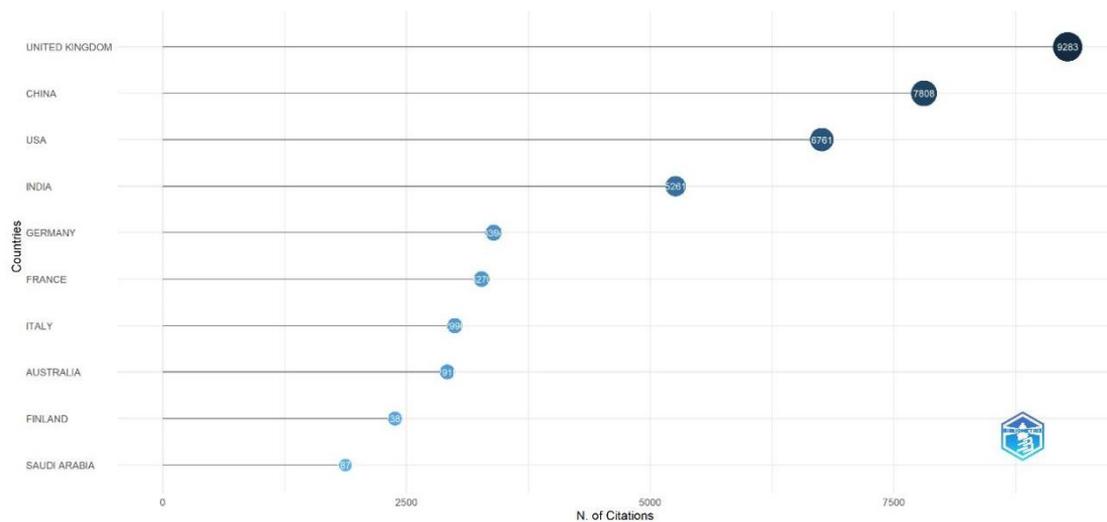

*Source: Created by authors*

### 4.7. Production of affiliations over time

*Figure 8: Affiliations production over time.*

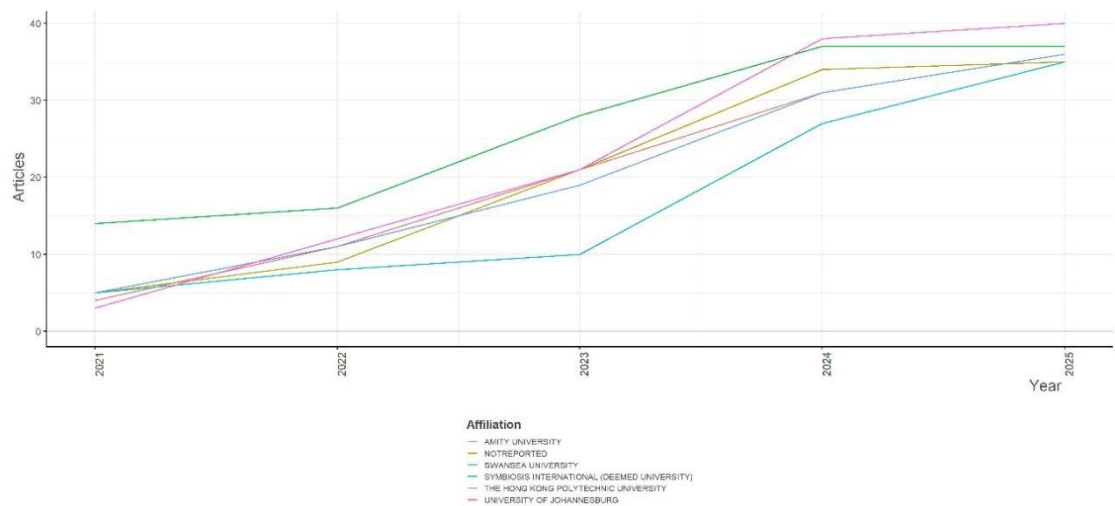

*Source: Created by authors*

Swansea University led in research paper production in the period of 2021-2024, achieving nearly 40 publications in 2024, before being surpassed by the University of Johannesburg in 2024-2025 reaching 40 publications. Unreported universities ranked the third in terms of affiliation production over time, reaching 30 publications in 2024, followed by Amity University, the Hong-Kong Polytechnical University and Symbiosis International (Deemed university).

**4.8. Words Frequency Over Time**

The frequency of words over time is an important parameter used in bibliometric analysis to identify the evolving trends and research focus in the scientific field. In Figure 8, the cumulative frequency of keywords from 2007 to 2023 is visualized. The term Artificial Intelligence showed a continuity in growth, in particular after 2018, exceeding all others by a large margin, reflecting its pivotal role in recent scientific discourse.

The terms Machine Learning, Deep Learning, and Neural Network also showed high growth, indicating a rising interest in these fields. Terms such as 'expert system' or 'fuzzy logic' are less dynamic, indicating their relative decline, stagnation, or progressive replacement in the scientific vocabulary by more recent or more expansive concepts.

*Figure 9: Word frequency over time.*

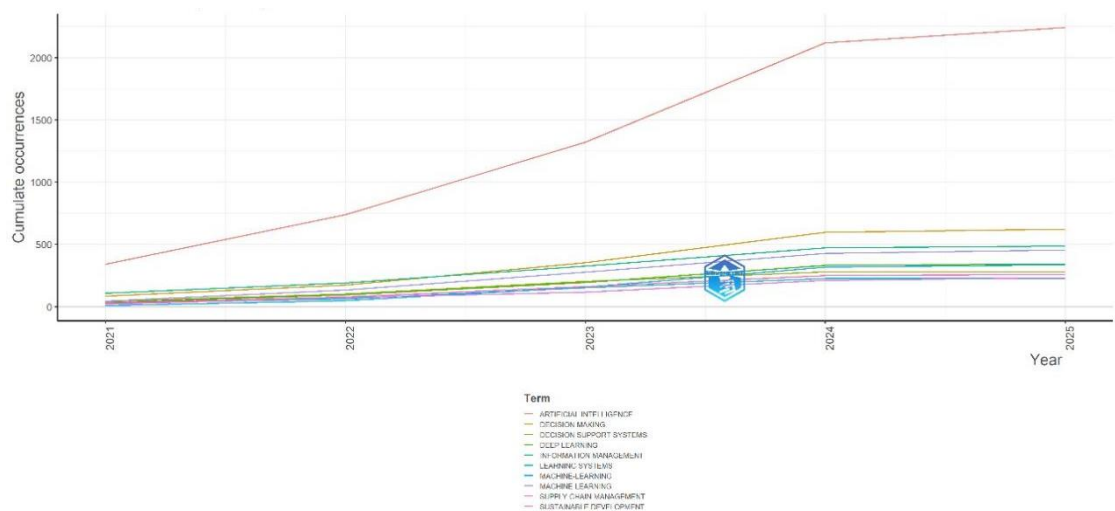

*Source : Created by authors*

**4.9. Sources' production over time**

The cumulative evolution of scholarly productivity from leading academic sources contextualizes past dynamics for the purpose of better interpreting current and emerging trends. This analysis allows for the identification of dominant journals to monitor and highlights potential shifts towards new sources or emerging areas of specialization in line with contemporary developments in research.

The journal named IEEE Transactions on Engineering Management is clearly emerged from the other sources by showing a constant and linear growth in publications, and achieving approximately 400 cumulative times by 2020. This dominance shows its central position in the present research status. Other sources including the Journal of Cleaner Production, European Journal of Control, and Sustainability illustrate moderate but solid growth indicating sustained contributions to their disciplines.

*Figure 10: Sources' production over time.*

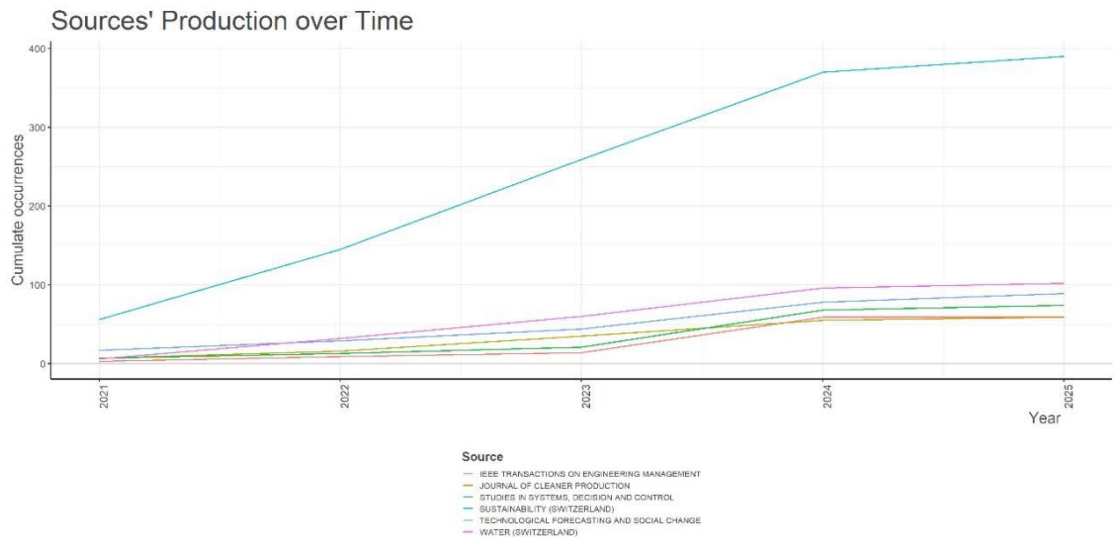

*Source : Created by authors*

### 4.10. Co-occurrence of keywords

In order to highlight the dominant concepts in the literature and to map the relationships between them, a color-coded cluster approach was used, Figure 10 highlights the thematic groupings using keyword frequencies and co-apparition in publications.

The term 'artificial intelligence' was shown to be the most frequent, therefore highlighted as the center of the themes network in figure 10, underscoring its thematic dominance, this central theme is associated with a high density of connections, highlighting its cross-disciplinary relevance.

*Figure 11: Co-occurrence of keywords.*

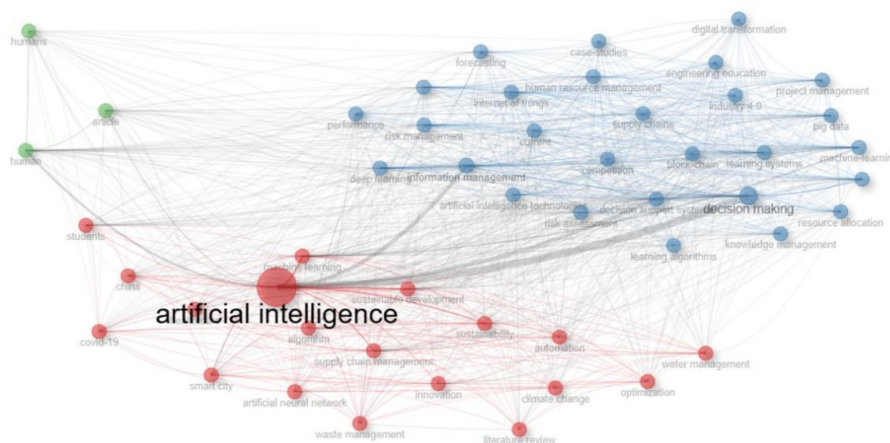

*Source: Created by authors*

The figure also highlights the red cluster, centered around the main theme, and including topics related to sustainability, urban planning, innovation, data sharing, waste management and COVID-19. the central position of the red cluster highlight the applications of AI in relationship to societal and environmental contexts, all based upon the cited literature in this study.

74

Moreover, a blue cluster appears in the figure illustrating, themes such as decision-making, digital transformation, project management, knowledge sharing and learning systems. Thus, underscoring the use of AI in such issues, notably for decision-making.

A third cluster less dominant, appears to highlight themes like humans and education, highlighting the decreased but present mention of human-machine interactions in the literature.

### 4.11. The thematic map

in order to provide a strategic overview of the output of science between 2021 and 2025, a thematic map was elaborated. Four quadrants were established, to reflect the dimensions of the degree of relevance (centrality) and the degree of them development (density).

The upper right quadrant, reflecting motor themes, illustrate topics such as decision-making, information management and ML. Being central and well-developed themes, those areas are characterized as mature and dynamic, therefore driving the evolution of science, especially in terms of connectivity with AI and data analysis.

Illustrated in the lower right quadrant, basic themes are found, with key words scu as AI, sustainability and innovation being highlighted. Those themes while being central to the scientific output is defined as low-density themes, therefore suggesting that they are in a phase of conceptual structuration or are in a phase of in-depth exploration.

Nich themes are represented in the upper right quadrant, including education, humans, and organization and management, while being well-developed areas, they are less central due to the fact that those areas are specialized or cross-sectoral, showing high internal coherence but limited connectivity.

The lower left quadrant, labeled emerging or declining themes, includes keywords like humans, article, and learning. Their low centrality and density indicate either emerging areas that are still underexplored or topics that are losing relevance in the current research landscape.

*Figure 12: Thematic map.*

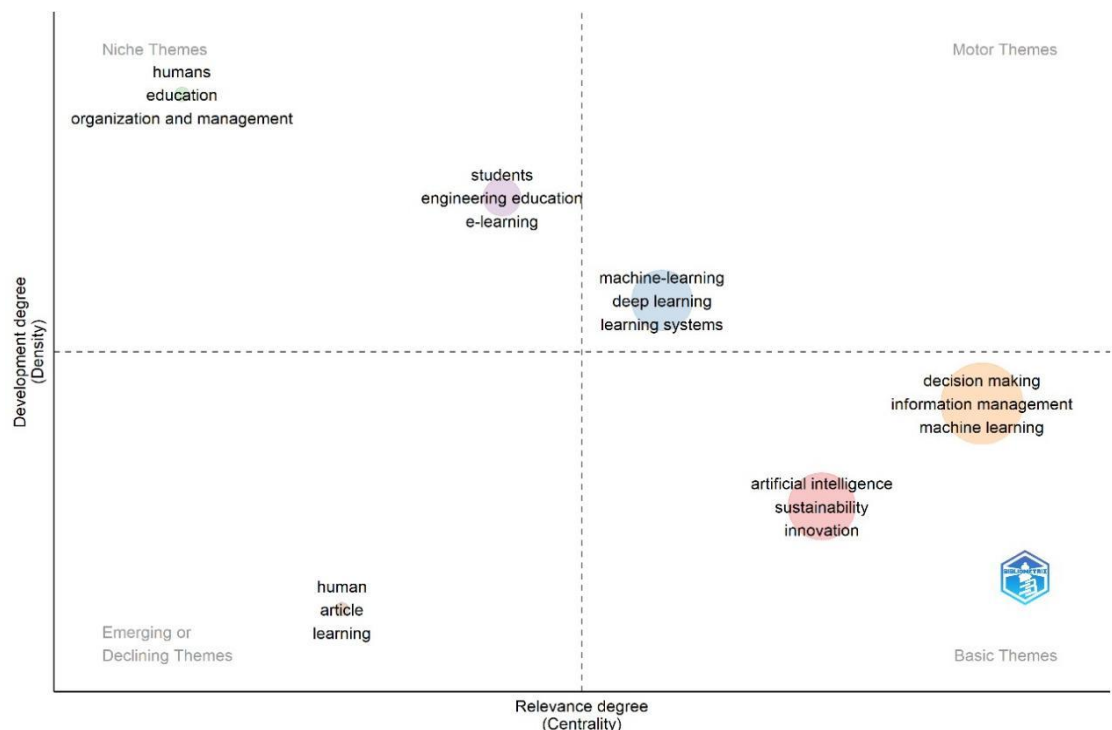

*Source: Created by authors*

### 4.12. Trending topics

Emerging and recurring themes in scientific production are highlighted in Figure 12 which indicate both duration of appearance and frequency of term occurrences. Such as software engineering, marketing, management, and water supply, already existing and trending in 2021 are suggested to be a continuation of preexisting research, but while somewhat old, their frequency is relatively modest, thus indicating a moderate scientific interest

ML, risk management, decision-making and AI, began to appear at higher frequencies starting 2022, progressively establishing a central dominance in terms of research axis and interest, in relation to challenges of digital transformation, uncertainty management, and the automation of decision-making processes.

By 2024, the emergence of terms like green development, libraries, and libraries- reflects a thematic shift towards environmental and informational concerns. This trend may indicate a realignment of scientific priorities in response to the challenges of sustainable development and knowledge management.

*Figure 13: Trending topics.*

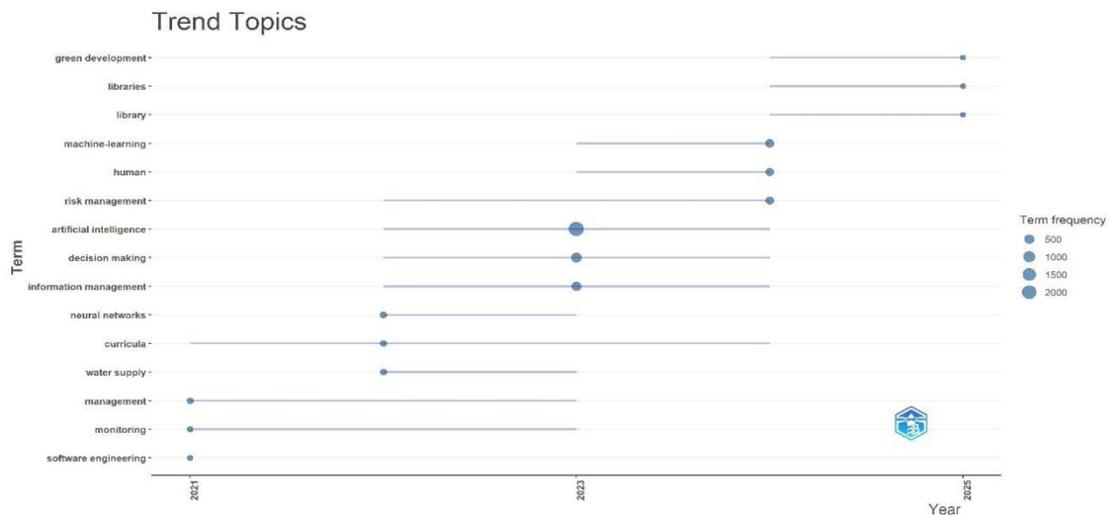

*Source: Created by authors*

### 4.13. Word cloud

The Figure 13 indicates the most frequently used terms in the scientific landscape over the 2021-2025 period. The term 'Artificial intelligence' features prominently, emphasizing its central and all-encompassing role in the present scientific production.

*Figure 14: Word cloud.*

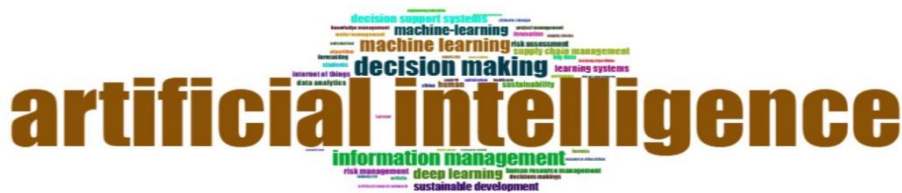

*Source: Created by authors*

Other keywords are appearing with high frequency, including machine learning, decision-making, and information management, which reflects a high interest in data analysis and decision-support tools. The occurrence of words like deep learning, sustainable development

76

and smart systems also demonstrate the expanding links between AI and the objectives of sustainability and systematic improvement.

## 5. Discussion

This bibliometric analysis provides a comprehensive overview of the evolving landscape of AI research within management studies from 2021 to 2025. The annual scientific production reveals dynamic shifts in scholarly interest and output, which can be interpreted through the lens of technological adoption and emerging research priorities in management science.

The data show a steady growth phase from 2021 to 2023, culminating in a peak during 2023–2024 with publication volumes exceeding 2000 papers. This rapid increase aligns with broader trends noted in prior bibliometric studies on AI, where heightened global interest corresponds with accelerated AI adoption across various management functions (Chen et al., 2022; Kaur & Sharma, 2023). However, the marked decline observed in 2024–2025, dropping to fewer than 500 publications, is intriguing and may reflect shifting research funding, the maturation of some research streams, or emerging saturation in certain AI topics, a phenomenon previously reported by Tang et al. (2024) in bibliometrics of emerging technologies, it is also important to note that publication, integrating the time taken to comprehensively generate data, analyze them, and publish was not taken into account in this study, therefore interpretation of this shift should be taken with a critical point of view.

The examination of the most relevant journals corroborates this trajectory, with sustainability (Switzerland) demonstrating exponential growth in publications from 2021 to 2024. Its slight slowdown from 2024 onwards could signal a pivot in research focus or journal thematic realignments, a common occurrence in evolving scientific fields (Eck & Waltman, 2014). Key other journals such as water (Switzerland), Technological Forecasting and Social Change, and IEEE Transactions on Engineering Management maintain consistent increases in output, highlighting their central roles in disseminating AI research with practical and policy implications in management.

The trends of author productivity indicate that influential scholars drive the discussion. Wang-y et al. Lead research output and citation impact in accordance with bibliometric studies that demonstrate how major contributors create new themes and link academic communities (Glänzel & Schoepflin, 1999). Author contribution maps enable the identification of influential knowledge leaders as well as new researchers who advance the field's development throughout time.

The global scientific production map verifies that the United States and China and India lead the research activities in Artificial Intelligence. The economic leadership of countries with strong research capabilities and funding allocation patterns results in leading scientific production output (Klein et al., 2019). The regions of Latin America, Africa and Central Asia show limited research participation because these regions lack equal research capacities and resource distribution which studies by Chinchilla-Rodriguez et al. (2024) have confirmed.

The analysis of corresponding authors' affiliations demonstrates high levels of international cooperation, especially between China, India and the USA. Research quality and impact improve through resource sharing and knowledge exchange within collaborative networks (Wagner & Leydesdorff, 2005). These countries maintain their leading position in AI innovation of management worldwide because they publish leading research independently and collaboratively with other nations.

Citation metrics provide both detailed and complex findings regarding the influence of research. Although China and India produce the largest number of publications, they rank behind the United Kingdom in terms of citations annually. This citation pattern shows that researchers should focus more on research quality and impact instead of publication numbers because this phenomenon appears frequently in bibliometric research (Bornmann & Daniel, 2008). The UK's

77

leadership position demonstrates its foundational and widespread contributions to subsequent research which extends its impact throughout global discussions beyond quantitative publication metrics.

Affiliation productivity helps institutions demonstrate their leadership role in research activities. Swansea University emerged as the second most productive institution from 2021 to 2024 while the University of Johannesburg followed it as a new hub for AI management research that extends past dominant research institutions. Research programs funded by institutions drive both output and impact as demonstrated by Sugimoto et al. (2017) in their studies.

The analysis of keyword frequency patterns together with co-occurrence data helps explain dominant research directions. The gradual introduction of Artificial Intelligence and its technologies into management research becomes apparent through the continuous increase of terms starting from 2018. The decrease of terms like "expert systems" and "fuzzy logic" indicates that researchers now prefer more advanced and scalable AI approaches which aligns with AI evolutionary research by Jiang et al. (2022).

The thematic analysis reveals distinct areas of research which center on Artificial Intelligence as an overarching interdisciplinary field linked to sustainability and urban planning and innovation and data management. The analysis shows how management research with AI applications extends its scope to tackle complex social and environmental problems, according to Wamba-Taguimdje et al. (2020). Decision-making, digital transformation, and knowledge sharing form another core cluster, underscoring AI's role in enhancing managerial processes and organizational learning (Chatterjee et al., 2021). Research on humans and education remains less dominant, which indicates ongoing investigations of human-AI interaction and workforce transformation but at a more specialized level.

The research field shows its capacity to adapt to worldwide challenges and technological transformations through trending topics which emerged from 2021 to 2025, according to Bug hin et al. (2018). Research in software engineering and marketing and water supply management initially held precedence before researchers started focusing on developing green initiatives and knowledge management systems which reflects their rising interest in sustainability and information systems. The research aligns with management scholarship that demands unified solutions between artificial intelligence and sustainable development goals (Liengpunsakul, 2021).

The visualization shows AI-related terms dominate scientific discourse while sustainable development and smart systems receive growing attention. The field demonstrates maturity because it implements technological advancements within environmental and societal parameters according to Dwivedi et al. (2021) who advocates responsible AI deployment in management.

## 6. Conclusion

This bibliometric study provides a clear overview of the evolution of research on artificial intelligence in the field of management between 2021 and 2025. Despite steady growth until 2024, a recent decline in output suggests a shift in pace or priorities. The results reveal a strong geographical concentration of publications (notably in China, India, and the United States), along with a notable citation impact from the United Kingdom. The most prolific journals are primarily focused on technology management and sustainable development.

Thematically, research has gradually incorporated more complex and societal dimensions, moving beyond purely technical aspects of AI. Trends indicate a rise in topics related to sustainability, digital transformation, and automated decision-making.

In conclusion, research on AI in management is evolving toward a more integrated approach, where technological innovation, social responsibility, and organizational performance converge. These dynamics pave the way for new research questions in the coming years.